\documentclass[twocolumn,english,aps,prl,reprint,floatfix,notitlepage,footinbib,preprintnumbers]{revtex4-1}
\pdfoutput=1
\usepackage{lmodern}

\usepackage[T1]{fontenc}
\usepackage[latin9]{inputenc}
\usepackage{geometry}
\usepackage{color}
\usepackage{dsfont}
\usepackage{slashed}
\usepackage{ragged2e}
\usepackage{comment}
\usepackage{babel}
\usepackage{amsmath,amssymb,amsthm,enumitem}
\usepackage{graphicx}
\usepackage[usestackEOL]{stackengine}
\usepackage{esint}
\usepackage[framemethod=TikZ]{mdframed}
\usepackage[unicode=true,pdfusetitle,
 bookmarks=true,bookmarksnumbered=false,bookmarksopen=false,
 breaklinks=false,pdfborder={0 0 1},backref=false,colorlinks=true]
 {hyperref}
\hypersetup{pdftitle={Amplitudes and the Riemann Zeta Function},
 pdfauthor={Grant N. Remmen},
 citecolor=black,linkcolor=black,urlcolor=black}
\usepackage{breakurl}
\usepackage[hang,flushmargin]{footmisc} 
\allowdisplaybreaks
\makeatletter

\usepackage{etoolbox}
\apptocmd{\thebibliography}{\justifying\setlength{\leftskip}{7.4mm}}{}{}

 \@ifundefined{textcolor}{}
 {%
   \definecolor{BLACK}{gray}{0}
   \definecolor{WHITE}{gray}{1}
   \definecolor{RED}{rgb}{1,0,0}
   \definecolor{GREEN}{rgb}{0,1,0}
   \definecolor{BLUE}{rgb}{0,0,1}
   \definecolor{CYAN}{cmyk}{1,0,0,0}
   \definecolor{MAGENTA}{cmyk}{0,1,0,0}
   \definecolor{YELLOW}{cmyk}{0,0,1,0}
 }

\usepackage{babel}

\makeatletter
\def\simgt{\mathrel{\lower2.5pt\vbox{\lineskip=0pt\baselineskip=0pt
           \hbox{$>$}\hbox{$\sim$}}}}
\def\simlt{\mathrel{\lower2.5pt\vbox{\lineskip=0pt\baselineskip=0pt
           \hbox{$<$}\hbox{$\sim$}}}}
\makeatother

\newcommand{\be}{\begin{equation}}
\newcommand{\ee}{\end{equation}}
\newcommand{\Fig}[1]{Fig.~\ref{#1}}
\newcommand{\Eq}[1]{Eq.~(\ref{#1})}

\begin{document}

\pagestyle{plain}

\title{Amplitudes and the Riemann Zeta Function}

\author{Grant N. Remmen}
\affiliation{Kavli Institute for Theoretical Physics}
\affiliation{Department of Physics, University of California, Santa Barbara, CA 93106}
\thanks{\vspace{-4mm}e-mail: \url{remmen@kitp.ucsb.edu}}

%%%%%%%%%%%%%%%%%%%%%%%%%%%%%%%%%%%%%%%%%%%%%%%%%%%%%%%%%%%%%%%%%%%%%%%%%%%%

\begin{abstract}
\noindent 
Physical properties of scattering amplitudes are mapped to the Riemann zeta function.
Specifically, a closed-form amplitude is constructed, describing the tree-level exchange of a tower with masses $m_n^2 = \mu_n^2$, where $\zeta\left(\frac{1}{2} \pm i\mu_n\right)  = 0$.
Requiring real masses corresponds to the Riemann hypothesis, locality of the amplitude to meromorphicity of the zeta function, and universal coupling between massive and massless states to simplicity of the zeros of $\zeta$.
Unitarity bounds from dispersion relations for the forward amplitude translate to positivity of the odd moments of the sequence of $1/\mu_n^2$.
\end{abstract}
\maketitle

\noindent{\bf Introduction.}---The Riemann zeta function, an object of central interest in number theory, is defined as
\be
\zeta(z) = \sum_{n=1}^\infty \frac{1}{n^z}  = \prod_{p\;{\rm prime}} \frac{1}{1-p^{-z}}
\ee
for ${\rm Re}(z)>1$, and by analytic continuation to the rest of the complex plane.
The function is analytic everywhere except for a simple pole at $z=1$ corresponding to the divergent harmonic series.
Despite the importance of the zeta function in mathematics and physics, from number theory to path integrals, many questions remain. Of particular interest is the location of its zeros. While $\zeta(z)$ exhibits trivial zeros at negative even integers by the functional equation,
\be
\zeta(z) = 2^z \pi^{z-1} \sin(\pi z/2)\Gamma(1-z)\zeta(1-z),\label{eq:functional}
\ee
it also possesses infinitely more zeros.
The known examples of these nontrivial zeros, which number in the trillions~\cite{Platt}, all lie on the critical line, 
$\zeta\left(\frac{1}{2} \pm i \mu\right) = 0$ for $\mu$ real.
(Throughout, we take ${\rm Re}(\mu) >0$: $\mu_1 \simeq 14.135$, $\mu_2 \simeq 21.022$, etc.)
The conjecture that $\mu$ is real for all nontrivial zeros of $\zeta(z)$ is the Riemann hypothesis, one of the most celebrated and fundamental extant problems in mathematics, with important consequences for the asymptotic distribution of the primes~\cite{Goldstein}.
Other open questions include whether all nontrivial zeros are simple ones~\cite{Simple,Montgomery}, as well as statistical properties of the zeros and asymptotic behavior of $\zeta$ on the critical line.

In this Letter, we will connect properties of the zeta function, including the Riemann hypothesis, to scattering amplitudes.
The idea of relating mathematical properties of the zeta function to a physical system  dates back a century to the Hilbert-P\'olya  conjecture~\cite{Polya,Montgomery} that the $\mu_n$ correspond to the eigenvalues of some quantum mechanical Hamiltonian.
Much work has been done to attempt to find such an operator (see, e.g., Refs.~\cite{BerryKeating,Srednicki:2011zz,Bender:2016wob}) or to identify other connections to physics, including Dyson's observation of the relation between the two-point function of the Gaussian unitary ensemble and Montgomery's pair correlation conjecture~\cite{Montgomery}, as well as interpretations of the phase of $\zeta$ in quantum chaotic nonrelativistic scattering~\cite{Gutzwiller,Bhaduri:1994de}.

Despite this progress, however, there has been relatively little work on the zeta function in the context of relativistic scattering amplitudes.
The program of reinterpreting a compelling mathematical object as an amplitude---before a Hamiltonian is found, and as a guide toward developing new and interesting physics---has notable precedent. 
In casting Euler's beta function as a scattering amplitude, Veneziano's expression~\cite{Veneziano}, and the search for a system producing this amplitude, led to the development of string theory.
In fact, as shown by Freund and Witten~\cite{FreundWitten}, the Veneziano amplitude itself can be written in terms of ratios of Riemann zeta functions, but in such a way that the nontrivial zeros cancel out~\cite{He:2015jla}.
This leaves open the question of a scattering amplitude whose structure depends on the nontrivial zeros of $\zeta$.

This is the question that will be answered in this work. In particular, we will build a scattering amplitude ${\cal M}(s,t)$, written in compact, closed form, for which there is an elegant correspondence between various physical properties and (known or conjectured) attributes of the Riemann zeta function:
\begin{widetext}
\begin{center}
\begingroup
\renewcommand{\arraystretch}{1}
\begin{tabular}{c c c}
${\cal M}(s,t)$ & $\big\vert$ & $\zeta(z)$\\
\hline
Poles at $s,u=m_n^2$ for $m_n$ real & $\longleftrightarrow$& Riemann hypothesis \\
Locality (simple poles) & $\longleftrightarrow$ & Meromorphicity\\
Universal coupling & $\qquad\longleftrightarrow\qquad$& Simple zero conjecture \\ Dispersive bounds from analyticity/unitarity
& $\longleftrightarrow$ & Positive odd moments of $\mu_n^{-2}$ sequence\\
On-shell constructibility & $\longleftrightarrow$ & Hadamard product expansion \\
CPT invariance & $\longleftrightarrow$ & Reflection of zeros across critical line
\end{tabular}
\endgroup
\end{center}
\end{widetext}
The existence of such an amplitude reframes the Hilbert-P\'olya problem and suggests that seeking a theory that naturally reproduces the form of our ${\cal M}(s,t)$ could lead to a solution of this conjecture, as well as interesting physical insights in their own right.

The purpose of this Letter will therefore be to construct such an amplitude ${\cal M}(s,t)$  and---without proving the Riemann hypothesis---map out the relations between its properties and features of the zeta function summarized above.
This Letter is structured as follows. 
First we construct an amplitude with the desired properties and study its analytic structure and uniqueness. 
We next examine the forward limit of our amplitude in the context of analytic dispersion relations, both as a verification of its analytic and asymptotic structure and as a means of physically deriving identities involving the nontrivial zeros of $\zeta$. 
After proving that the amplitude is on-shell constructible and relating this construction to the product form of the zeta function, we explore connections of the symmetry of the zeros across the critical line with CPT symmetry and comment on potential future directions.
\newline

\begin{figure}[t]
\begin{center}
\includegraphics[width=\columnwidth]{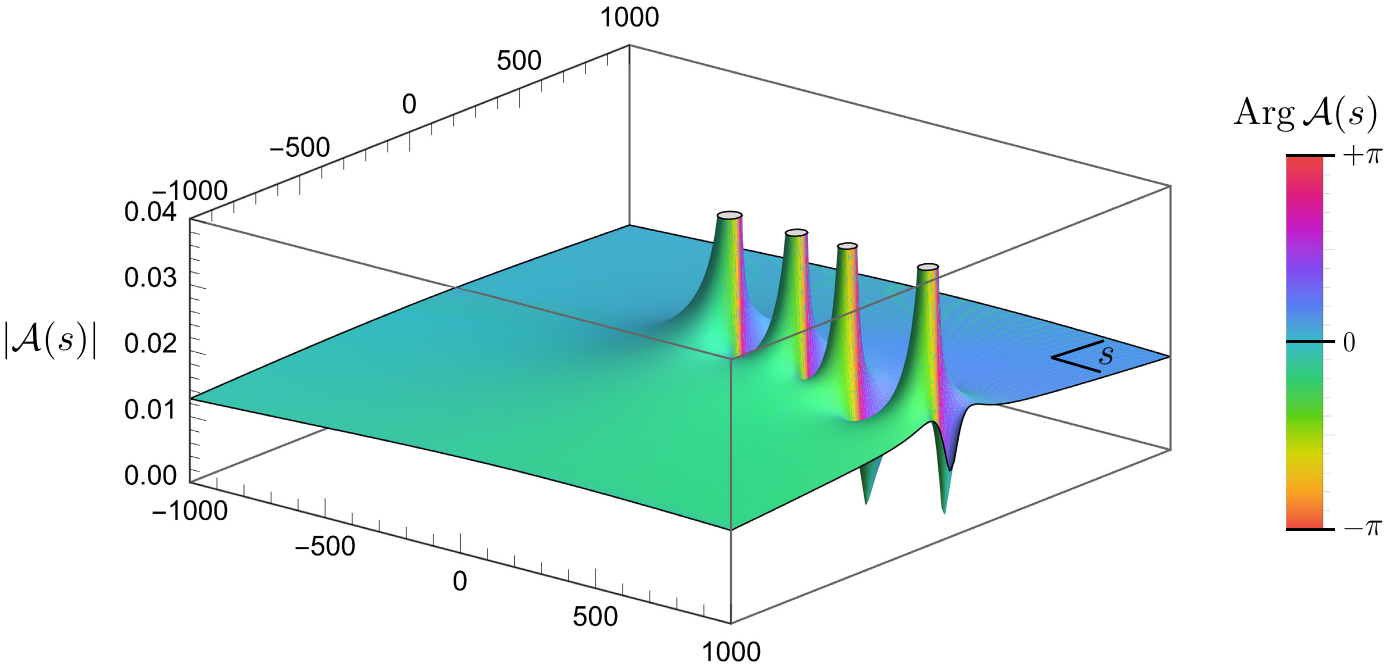}
\end{center}\vspace{-4mm}
\caption{Illustration of ${\cal A}(s)$ defined in Eq.~\eqref{eq:A}. As shown in text, ${\cal A}(s)$ is meromorphic, with poles at $s=\mu_n^2$ corresponding to the nontrivial zeros of the Riemann zeta function, $\zeta\left(\frac{1}{2}\pm i \mu_n\right) =0$.
}
\label{fig:3D}
\end{figure}

\noindent{\bf Building the amplitude.}---Let us define a function of a single complex variable $s$,
\be 
\begin{aligned}
{\cal A}(s)  &=  -\frac{i}{4\sqrt{s}}\left[\psi\left(\tfrac{1}{4} + \tfrac{i}{2}\sqrt{s}\right) + \frac{2 \zeta'\left(\frac{1}{2} + i\sqrt{s}\right)}{\zeta\left(\frac{1}{2}+ i\sqrt{s}\right)}\right] \\& \hspace{4mm} +\frac{i\log\pi}{4\sqrt{s}} - \frac{1}{s+\frac{1}{4}},
\end{aligned}\label{eq:A}
\ee
where $\zeta'(z) = {\rm d}\zeta(z)/{\rm d} z$ and $\psi(z)$ is the digamma function; see Fig.~\ref{fig:3D}. In terms of the Landau-Riemann (capital) xi function $\Xi(z)=\xi\left(\frac{1}{2}+iz\right)$, where $\xi(z)$ is defined as $\frac{1}{2}z(z-1)\pi^{-z/2}\Gamma\left(\frac{z}{2}\right)\zeta(z)$, application of the functional relation~\eqref{eq:functional} allows us to write ${\cal A}(s)$ very compactly:
\be 
{\cal A}(s) = -\frac{{\rm d} \log\Xi(\sqrt{s})}{{\rm d}s}.\label{eq:Acompact}
\ee
We then use ${\cal A}$ to define an amplitude describing the four-point scattering of massless particles in terms of the Mandelstam variables, $s=-(p_1 + p_2)^2$, $t=-(p_1 + p_3)^2$, and $u=-s-t$~\footnote{Throughout, we will use unitless Mandelstam variables. To restore units, one can send $s\rightarrow s/\Lambda^2$, etc., everywhere, with $\Lambda$ defining the scale of the ultraviolet completion. Similarly, the amplitude  can be rescaled by $\kappa^2$ for real coupling constant $\kappa$.}:
\be
{\cal M}(s,t) = {\cal A}(s) + {\cal A}(u).\label{eq:M}
\ee
As we will see, unlike an arbitrary complex function, ${\cal M}(s,t)$ possesses all of the standard properties of a scattering amplitude---unitarity, analyticity, and locality---and describes the tree-level exchange of heavy states in the $s$ and $u$ channels, with spectrum $m_n = \mu_n$.

Let us see how these properties come about from our definition of ${\cal A}(s)$.
One can show that $\Xi(z)$ is everywhere analytic (i.e., entire), with roots corresponding to the zeros of the zeta function, $\Xi(\mu_n)=0$, and that the functional equation~\eqref{eq:functional} becomes $\Xi(z)=\Xi(-z)$.
As a result, even though $\sqrt{s}$ has a branch cut at real $s<0$, $\Xi(\sqrt{s})$ is entire~\cite{Titchmarsh}, and so ${\cal A}(s)$ from \Eq{eq:Acompact} is meromorphic, with simple poles at $s=\mu_n^2$.

More explicitly, starting from \Eq{eq:A}, direct evaluation of this limit for ${\cal A}$ yields $\lim_{\epsilon\rightarrow 0} {\cal A}(s+i\epsilon) - {\cal A}(s-i\epsilon) = 0$ for real $s$, using the reality of $\zeta$ and $\psi$ on the real line.
Since the digamma and zeta functions are meromorphic, we therefore have that ${\cal A}(s)$ is meromorphic as well, i.e., ${\cal M}(s,t)$ is analytic except for poles. 
It thus remains to examine the behavior of ${\cal A}$ near the poles/zeros of $\psi$ and $\zeta$.
The arguments of the digamma function and zeta function in Eq.~\eqref{eq:A} are chosen such that the simple zeros of $\zeta(z)$ at $z=-2n$ coincide with the simple poles of $\psi(z)$ at $z=-n$, at $s=-(4n+1)^2/4$ for positive integer $n$, with the result that the poles cancel and ${\cal A}(s)$ is analytic there. 
The only other digamma pole, $\psi(0)$, occurs at $s=-1/4$, which is also the location of the pole in the $1/(s+\frac{1}{4})$ term; explicit evaluation shows that $\lim_{s\rightarrow -1/4} {\cal A}(s)$ is finite and equal to $(2+\gamma-\log 4\pi)/2$, where $\gamma$ is the Euler-Mascheroni constant.
Finally, the $s\rightarrow 0$ limit is finite:
\be
\begin{aligned}
\frac{c_0}{2} = \lim_{s\rightarrow 0} {\cal A}(s) &=  -4+\frac{\pi^{2}}{8}+G+\frac{\zeta''\left(\frac{1}{2}\right)}{2\zeta\left(\frac{1}{2}\right)} \\&\hspace{3.9mm}-\frac{1}{8}\left(\gamma+\frac{\pi}{2}+\log8\pi\right)^{2},
\end{aligned}\label{eq:c0}
\ee
writing $G$ for Catalan's constant $\sum_{k=0}^\infty (-1)^k/(2k+1)^2$ and defining $c_0 = {\cal M}(0,0)$.

Using the Hadamard product form for the zeta function, along with differentiation of the functional equation~\eqref{eq:functional} (which allows for computation of odd derivatives of $\zeta$ at $1/2$) and various gamma function relations, one can find an identity for $\zeta''(1/2)$ in terms of a sum over the nontrivial zeros of $\zeta$, which yields the beautiful result:
\be
c_0 = \sum_{n=1}^\infty \frac{2}{\mu_n^2} \simeq 4.6210\times 10^{-2}.\label{eq:c0sum}
\ee
As we will see, sums of this form, containing powers of the sequence $1/\mu_n^2$, will have important connections to analytic dispersion relations.

The only remaining candidate poles in ${\cal A}(s)$ correspond to the nontrivial zeros of the zeta function, $\zeta\left(\frac{1}{2}+ i\mu_n\right)=0$ at $s=\mu_n^2$.
If the Riemann hypothesis is true, then all of these poles occur on the real $s$ axis. Further, we find that each positive-$s$ residue satisfies $\oint_{s=\mu_{n}^{2}}i{\cal A}(s){\rm d}s>0$, as required by unitarity for a physical pole in an amplitude, i.e., if we move each pole at $s=\mu_{n}^{2}$ to $s=\mu_{n}^{2}-i\epsilon$ in Feynman's $i\epsilon$ formalism, then we have ${\rm Im}\,{\cal A}(s)>0$. 
Specifically, if the $n$th nontrivial zero of $\zeta$ has order $g_n$, i.e., $\zeta(z)\sim (z-z_n)^{g_n}$ near $z_n$, then $\lim_{s\rightarrow \mu_{n}^{2}}(\mu_{n}^{2}-s){\cal A}(s)=g_n$,  so 
\be 
\oint_{s=\mu_{n}^{2}}i{\cal A}(s){\rm d}s=2\pi g_n.\label{eq:oint}
\ee 
Our amplitude behaves as a tower of tree-level exchanges, with spectrum of masses $m_{n}$ in one-to-one correspondence with the nontrivial zeros of the Riemann zeta function,
\be 
m_{n}=\mu_{n}.
\ee 
For a theory with scattering described by ${\cal M}(s,t)$, the Riemann hypothesis then becomes the physical requirement of real masses for the on-shell states in the spectrum.

If all of the zeros of the Riemann zeta function are simple as has been conjectured~\cite{Simple,Montgomery}, then $g_n = 1$ for all $n$, in which case the massive states in the amplitude enjoy a universal coupling to the scattering states; if not, then the couplings are controlled by the order $g_n$.
We can parameterize $g_n \neq 1$ by allowing multiple redundant $\mu_n$ in any sum or product (e.g., \Eq{eq:c0sum}), which we will do henceforth.
Our amplitude exhibits locality, i.e., near each pole, ${\cal A}(s) \sim 1/(-s+\mu_n^2)$.
A failure of locality in ${\cal A}(s)$ via a pole $\sim 1/(-s+\mu_n^2)^k$ for some $k>1$ would require $\zeta(z)\sim \exp[\alpha/(z-z_n)^{k-1}]$ near the corresponding zero $z_n$, for some $\alpha$. This would be an essential singularity: depending on the direction of approach, $\zeta$ could go to zero or infinity as $z\rightarrow z_n$.
Hence, locality in ${\cal A}(s)$ is enforced by the fact that the zeta function is meromorphic and therefore lacks essential singularities.

Before exploring other interesting properties of ${\cal M}$, we first argue that this is the simplest candidate amplitude satisfying the following requirements: {\it i.})~${\cal M}$ is analytic everywhere except poles corresponding to the nontrivial zeros of the Riemann zeta function, and these poles are real if the Riemann hypothesis holds; {\it ii}.)~each pole has positive residue as in Eq.~\eqref{eq:oint}; and {\it iii}.)~the forward amplitude satisfies $\frac{{\rm d}^2}{{\rm d}s^2} {\cal M}(s,0) \neq 0$ in the limit $s\rightarrow 0$.
Taking the ansatz that ${\cal M}$ is separable into channels ${\cal A}(s)$ and ${\cal A}(u)$ is a natural choice that enforces crossing symmetry. To satisfy condition {\it i}.) on the nontrivial zeros, one could take ${\cal A}(s) \sim 1/\zeta\left(\frac{1}{2}+i s\right)$.
However, this choice runs afoul of requirement {\it ii}.), which can be simply corrected by multiplying by the derivative of the zeta function, which guarantees that each pole at a nontrivial zero has a residue of the same sign.
Canceling the trivial zeros in $\zeta$ and the pole at $z=1$ necessitates adding an infinite tower of other terms, which result in digamma and algebraic terms as in \Eq{eq:A}.
Finally, the radicals in \Eq{eq:A} are necessary, since if we take the forward amplitude and send $s\rightarrow s^2$ to eliminate the square roots, we find that ${\cal M}(s^2,0)-{\cal M}(0,0) \propto s^4$ near $s=0$; this is too soft to satisfy condition {\it iii}.)---which as we will discuss in the next section comes from dispersion relation bounds~\cite{Adams:2006sv} (cf.~the Galileon~\cite{Nicolis:2009qm})---so we resolve this problem by introducing $\sqrt{s}$, resulting in ${\cal A}(s)$ as given in \Eq{eq:A}.
Hence, our form for ${\cal M}(s,t)$ is arguably the simplest possible amplitude relevant to the Riemann hypothesis, up to adding or multiplying by an entire function.
\linebreak

\noindent{\bf Analytic dispersion relations.}---Forward amplitudes in an infrared effective field theory coming from a well behaved ultraviolet completion are known to possess positivity properties coming from analytic dispersion relations. 
In particular, if ${\cal M}(s,t)$ is indeed an amplitude, we should find that
\be 
\lim_{s\rightarrow 0}\frac{{\rm d}^{2k}}{{\rm d}s^{2k}}{\cal M}(s,0)>0
\ee
for all $k>0$.
This is a classic consequence of analyticity and unitarity~\cite{Adams:2006sv}. 

Computing a contour integral 
\be 
c_{2k} = \frac{1}{2\pi i} \oint_{\cal C} \frac{{\rm d} s}{s^{2k+1}} {\cal M}(s,0)\label{eq:ck}
\ee
for ${\cal C}$ a small contour around the origin, analyticity of ${\cal M}$ allows ${\cal C}$ to be deformed to a new contour running just above and below the real $s$ axis, plus a circle at infinity.  We note that the definitions of $c_0$ match in Eqs.~\eqref{eq:c0} and \eqref{eq:ck}.
The optical theorem (i.e., unitarity) and crossing symmetry imply that $c_{2k} = \frac{2}{\pi} \int_0^\infty \frac{{\rm d}s}{s^{2k}}\sigma(s) + c^{(2k)}_\infty$, where $\sigma(s)$ is the (positive) cross section associated with the scattering in the amplitude and $c^{(2k)}_\infty=\frac{1}{2\pi i}\oint_{|s|=\infty} \frac{{\rm d} s}{s^{2k+1}} {\cal M}(s,0)$ is a boundary term.
A nonzero boundary term for some $k \geq 0$ would imply that $\Xi(z)$ grows at least as fast as $\exp(\alpha z^{4k+2})$ for some $\alpha$ (i.e., growth order at least $4k+2$), which is inconsistent with the fact that $\Xi(z)$ has a known growth order of unity~\cite{Titchmarsh}; thus, all of the $c^{(2k)}_{\infty}$ must vanish.
For example, since we have already shown that ${\cal M}(s,0)$ is analytic everywhere in the complex $s$ plane except for the poles at $s=\pm \mu_n^2$, using the value of the residue at each pole in Eq.~\eqref{eq:oint} we find that
\be 
c_0 \! = c_\infty^{(0)} + \frac{1}{2\pi i}\sum_n\! \oint_{s = \mu_n^2} \!\!\! \frac{{\rm d}s}{s} {\cal M}(s,\! 0) = c_\infty^{(0)} + \sum_n \frac{2}{\mu_n^2},\label{eq:c0analytic}
\ee
so \Eq{eq:c0sum} implies that $c_\infty^{(0)} = 0$.
The coefficients in \Eq{eq:ck} define an expansion ${\cal M}(s,0) = \sum_{k=0}^\infty c_{2k}s^{2k}$ near $s=0$.
Analogously with \Eq{eq:c0analytic}, we have the explicit prediction for the value of $c_{2k}$:
\be 
c_{2k}=\sum_{n=1}^{\infty}\frac{2}{\mu_{n}^{2(2k+1)}}.\label{eq:pred}
\ee

\begin{figure}[t]
\begin{center}
\includegraphics[width=\columnwidth]{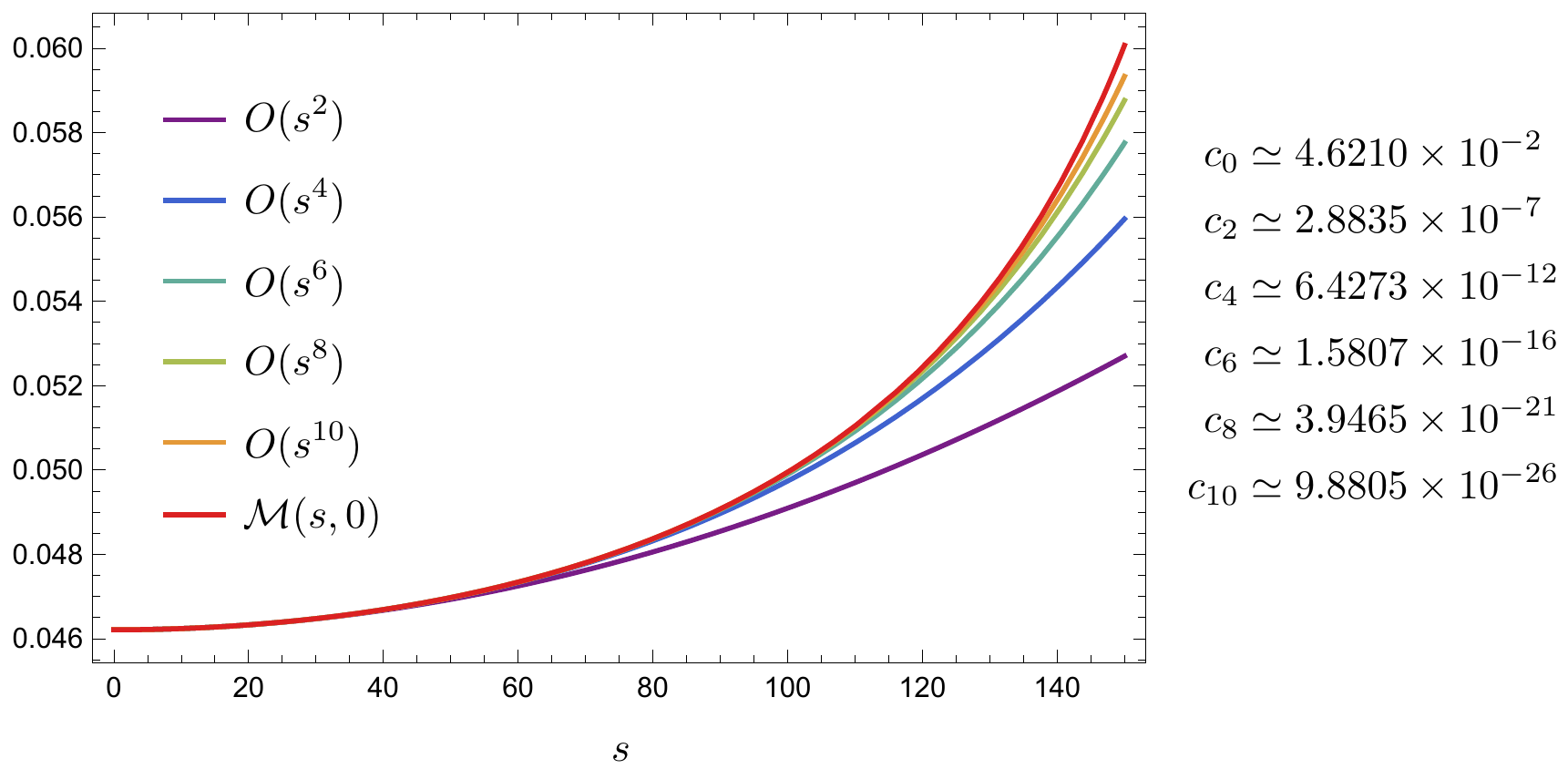}
\end{center}\vspace{-5mm}
\caption{Taylor series of ${\cal M}(s,0)$ near $s=0$. All even derivatives of ${\cal M}(s,0)$ at $s=0$ are positive, as required by analyticity/unitarity for a forward amplitude or alternatively by the Riemann hypothesis.
}
\label{fig:taylor}
\end{figure}

The Riemann hypothesis would imply the positivity of the $c_{2k}$ required by unitarity and analyticity.
See \Fig{fig:taylor} for an illustration.
This is a nontrivial check of the analytic and asymptotic structure of ${\cal M}(s,0)$, confirming that it indeed behaves like a forward amplitude.
Our amplitude construction allows for the derivation of further remarkable identities akin to the $c_0$ relation in \Eq{eq:c0sum}.
Defining the normalized $n$th derivative $\zeta_{n}(z)=\zeta^{(n)}(z)/\zeta(z)$ and $\zeta_{n}^{k}(z)=\left[\zeta_{n}(z)\right]^{k}$, we have:
\be
\begin{aligned}
c_{2} &=\frac{1}{2}\lim_{s\rightarrow0}\frac{{\rm d}^{2}}{{\rm d}s^{2}}{\cal M}(s,0)\\
& =-128+ \frac{1}{7680}\psi^{(5)}\left(\tfrac{1}{4}\right)-\zeta_{1}^{6}\left(\tfrac{1}{2}\right) \\& \hspace{4mm} +3\zeta_{1}^{4}\left(\tfrac{1}{2}\right)\zeta_{2}\left(\tfrac{1}{2}\right)-\frac{9}{4}\zeta_{1}^{2}\left(\tfrac{1}{2}\right)\zeta_{2}^{2}\left(\tfrac{1}{2}\right)\\
 & \hspace{4mm} +\frac{1}{4}\zeta_{2}^{3}\left(\tfrac{1}{2}\right)-\zeta_{1}^{3}\left(\tfrac{1}{2}\right)\zeta_{3}\left(\tfrac{1}{2}\right) \\& \hspace{4mm}  +\zeta_{1}\left(\tfrac{1}{2}\right)\zeta_{2}\left(\tfrac{1}{2}\right)\zeta_{3}\left(\tfrac{1}{2}\right)-\frac{1}{12}\zeta_{3}^{2}\left(\tfrac{1}{2}\right)\\
 & \hspace{4mm} +\frac{1}{4}\zeta_{1}^{2}\left(\tfrac{1}{2}\right)\zeta_{4}\left(\tfrac{1}{2}\right)-\frac{1}{8}\zeta_{2}\left(\tfrac{1}{2}\right)\zeta_{4}\left(\tfrac{1}{2}\right) \\& \hspace{4mm} -\frac{1}{20}\zeta_{1}\left(\tfrac{1}{2}\right)\zeta_{5}\left(\tfrac{1}{2}\right)+\frac{1}{120}\zeta_{6}\left(\tfrac{1}{2}\right) \\
 &= \sum_{n=1}^\infty \frac{2}{\mu_n^6}.\label{eq:c2}
\end{aligned}
\ee
Like \Eq{eq:c0sum}, \Eq{eq:c2} can be proven exactly albeit laboriously, without appeal to our amplitude, using repeated differentiation of the functional equation and the  Hadamard product form of the zeta function, as well as various polygamma identities; the same should hold for all other $c_{2k}$.
As a check, the prediction in \Eq{eq:pred} can be confirmed to within a relative precision of one part in $10^{30}$ for $k=2,3,4,5$ by summing over numerical values of the zeros $\mu_n$ given in Ref.~\cite{lmfdb}.
While each order in \Eq{eq:pred} can be checked mathematically, what is remarkable is that our amplitude construction allows for much simpler, physical derivations of these identities.
\linebreak

\noindent{\bf On-shell constructibility.}---Given the properties we have found for ${\cal A}(s)$, our amplitude ${\cal M}(s,t)$ describes two massless scalars exchanging a tower of massive states in the $s$ and $u$ channels with constant, momentum-independent coupling.
For example, we could have two species of scalars, $\phi_1$ and $\phi_2$, scattering via $\phi_1\phi_2\rightarrow\phi_1\phi_2$. 
Alternatively, we could have instead defined ${\cal M}$ in \Eq{eq:M} with full Bose symmetry as ${\cal A}(s)+{\cal A}(t)+ {\cal A}(u)$ to describe the four-point scattering of a single scalar.
If there is a coupling $\propto\phi_1\phi_2 X$, where $X$ is a tower of states with masses $m_{n}^2=\mu_{n}^2$, then the tree-level amplitude for this theory will match Eq.~\eqref{eq:M}.
That is, our amplitude ${\cal M}(s,t)$ is on-shell constructible~\cite{Cohen:2010mi} from the three-point $\phi_1\phi_2 X$ amplitudes, which are all a constant (and universal for all $X$ if the simple zero conjecture holds).
The function defined in \Eq{eq:A} is equivalent to
\be 
{\cal A}(s) = \sum_n \frac{1}{-s+\mu_{n}^{2}},\label{eq:poles}
\ee
and hence
\be 
{\cal M}(s,t)=\sum_{n}\left(\frac{1}{-s+\mu_{n}^{2}-i\epsilon}+\frac{1}{-u+\mu_{n}^{2}-i\epsilon}\right).\label{eq:polesM}
\ee

This equality can be seen as follows. Define $\Delta(s)$ as the difference between the right-hand sides of Eqs.~\eqref{eq:A} and \eqref{eq:poles}.
As we have shown, since ${\cal A}(s)$ as defined in \Eq{eq:A} has poles only at $s=\mu_n^2$ with unit residue (writing any instance of multiple zeros as distinct $\mu_n$), it follows that $\Delta$ is entire.
Expanding in a Laurent series around $s=\infty$, the form of \Eq{eq:poles} at large $s$ and our previous result that \Eq{eq:A} possesses no pole at infinity together then imply that $\Delta$ is bounded, so by Liouville's theorem $\Delta$ is constant.
By the direct evaluation of ${\cal A}(0)$ in \Eq{eq:c0sum}, $\Delta(0)=0$, yielding \Eq{eq:poles}.
As a result of the form in \Eq{eq:polesM}, ${\cal M}$ will automatically satisfy the EFT-hedron constraints~\cite{Arkani-Hamed:2020blm}, beyond the dispersion relation bounds discussed above, which we expect would lead to streamlined derivations of more zeta function identities analogous to \Eq{eq:c2}.

Comparing \Eq{eq:poles} with the form of ${\cal A}$ in terms of $\Xi$ in \Eq{eq:Acompact}, we see that the on-shell constructible expression for the amplitude gives $\Xi(z) = \Xi(0)\, \prod_n [1 - (z^2/\mu_n^2)]$, the Hadamard product expansion of the xi function.
\linebreak

\noindent{\bf Discussion.}---The zeta function possesses various other properties that can be mapped to physical features of the scattering amplitude.
For example, its zeros are symmetric both across the real axis and across the critical line  at ${\rm Re}(z){=}1/2$ as a consequence of the Schwarz reflection principle $\zeta(\bar z) \,{=}\, \overline{\zeta(z)}$ and \Eq{eq:functional}. 
Hence, ${\rm Im}\,{\cal M}(s,0)$ is nonzero only by virtue of the $i\epsilon$ terms, going as a sum of $\pi \delta(\pm s{-}\mu_n^2)$.
This allows the optical theorem ${\rm Im}\,{\cal M}(s,0)\, {=}\, s\,\sigma(s)$ to respect momentum conservation, with nonzero $\sigma$ only when the external momenta  produce an on-shell intermediate massive state.
A consequence is that we can write the zeta zero-counting function $N(T)$---the number of $z$ for which $\zeta(z) = 0$ and $0\,{<}\,{\rm Im}(z)\,{\leq} \,T$---in terms of the cross section as $N(s_0^2) = \frac{1}{\pi}\int_0^{s_0} \sigma(s)\, {\rm d}s$.
Complex $\mu_n = M\,{-}\,iW$, violating the Riemann hypothesis, would contribute an additional imaginary part to the forward amplitude $\propto W$ for $W\,{\ll}\, M$.
Symmetry of zeros about the critical line ensures that such terms would come in pairs with $\pm W$, eliminating this extra contribution to ${\rm Im}\,{\cal M}(s,0)$.
As a resonance, these zeros represent a pair of decaying/growing modes, and the reflection of zeros about the critical line ensures that ${\cal M}$ obeys the CPT theorem.

Our construction of ${\cal M}(s,t)$ suggests various interesting generalizations. 
The construction of higher-point or loop amplitudes by gluing together copies of ${\cal M}$ merits investigation.
Moreover, while the momentum-independent coupling evident in \Eq{eq:poles} implies that the states exchanged in ${\cal M}$ are scalars, we could generalize ${\cal A}$ by introducing momentum dependence into the propagator numerators, thus encoding spin for the massive states.
Another compelling direction would be to construct the analogue of ${\cal A}$ from an arbitrary Dirichlet $L$-function, of which $\zeta$ is a special case.
Doing so would modify the spectrum, and the generalized Riemann hypothesis would relate reality of the masses and zeros on the critical line.
More broadly, replacing $\Xi(\sqrt{s})$ with an arbitrary entire function possessing real, positive zeros and the requisite boundary conditions could generalize the amplitude construction to other functions of mathematical interest.
Finally, the universality property of the zeta function~\cite{Voronin} and its consequences for the amplitude are worthy of study.
We leave such questions to future investigation.

\vspace{3mm}

\begin{acknowledgments}
\noindent {\it Acknowledgments:}
I thank Cliff Cheung, Yu-tin Huang, Pratik Rath, Nick Rodd, and Mark Srednicki for discussions and comments.
This work was supported at the Kavli Institute for Theoretical Physics by the Simons Foundation (Grant No.~216179) and the National Science Foundation (Grant No.~NSF PHY-1748958) and at the University of California, Santa Barbara by the Fundamental Physics Fellowship.
\end{acknowledgments}

\vspace{-5mm}

\bibliographystyle{utphys-modified}
\bibliography{ZetaAmplitudes}

\end{document}